# Variation of positiveness to enhance testing of specimens during an epidemic.

Usama Kadri[1]

[1]*School of Mathematics, Cardiff University, Cardiff CF24 4AG, UK*

**Rapid testing of appropriate specimens from patients suspected for a disease during an epidemic, such as the current Coronavirus outbreak, is of a great importance for the disease management and control. We propose a method to enhance processing large amounts of collected samples. The method is based on mixing samples in testing tubes in a specific configuration, as opposed to testing single samples in each tube, and accounting for natural virus amounts in infected patients from variation of positiveness in test tubes. To illustrate the efficiency of the suggested method we carry out numerical tests for actual scenarios under various tests. Applying the proposed method enhances the number of tests by order of magnitudes, where all positives are identified with no false negatives, and the effective testing time can be reduced drastically even when the uncertainty in the test is relatively high.**

The World Health Organization has declared the growing epidemic of novel coronavirus infectious disease (COVID-19) a global pandemic. The virus emerged in Wuhan, China, at the end of 2019, and as of April 22, 2020 over two and a half million cases were identified in 210 countries and territories, with nearly 180,000 deaths being reported[1]. In most countries around the world the number of cases is believed to be much larger than reported. The relatively low reported number is attributed to a number of factors, including but not limited to mismanagement of the epidemic at the political level, high ER visit costs, and a lack of resources that limit the number of tests drastically. For example in the UK 25,000 tests were carried out in the period since January 2020 and up until March 11, 2020, which is equivalent to the number of tests carried out in South Korea in two and half days according to the World Health Organization. Thus, while in some countries there is a (front-end) problem of sample collection, in other countries the main concern is in processing the collected samples (back-end problem). Here we are concerned with the back-end problem, namely processing a large amounts of collected samples.

Specimens can be collected from the upper respiratory tract as nasopharyngeal (through the nose) and oropharyngeal (through the mouth) swab or wash in an ambulatory regime[2]. The laboratory confirmation of the COVID-19 is based on the Nucleic Acid Amplification Tests (NAAT); the assay detects the genomic sequences of virus RNA by real-time reverse transcription polymerase chain reaction rRT-PCR.

The current outbreak has evoked researchers and experts from various fields to reevaluate the feasibility of multi-sample pools[3,4], where samples from a number of patients are mixed together, as opposed to testing individual samples. In this work, we propose an advanced testing method where samples from each patient are mixed in multiple tubes in a unique configuration, then variation of test "positiveness" of each tube are employed in order to calculate all possible positives. The first part of the method is by itself powerful when the percentage of infected patients is extremely low, as long as proper mixing is done – keeping in mind the dilution threshold (due to mixing) required for identifying the disease. However, as the percentage of the infected increases it becomes much more challenging to determine positives without performing new tests. To this end, an accurate quantitative approach (the second part of the proposed method) can be employed to determine all positives without risking having false negatives. For this part to be effective, an accurate method for quantification of PCR is required[5]. The proposed method takes into account the uncertainty (error) in the test. Even when the uncertainty increases (i.e. for less accurate tests), all positives are still obtained with no false negatives, though false positives start to arise as well. The mathematical model of the proposed method is presented in the following section.

## 1 Variation of positiveness: Mathematical model

Let $n$ be the number of patients (sample size), $m$ the size of the test tube set, and $l$ the size of a subset of the tube test set. Each patient sample is distributed to a different configuration of $l$ tubes. Thus, the maximum number of patients is given by

$$n = \frac{m!}{l!(m-l)!} \tag{1}$$

The test results in each tube $j = 1..l$ can be described by

$$R_j = \sum_{i=1}^{n} r_{ij} \delta_i, \quad (j = 1..l), \tag{2}$$

where $r_{ij}$ is the contribution of the $i$-th patient to the 'positiveness' of the $j$-th tube, and $\delta_{ij}$ is the delta function, being zero or unity for negatively or positively tested patients, respectively.

Thus, we can now construct a set of $n$ algebraic equations

$$\begin{pmatrix} r_1 & r_1 & r_1 & \ldots & \ldots & 0 & 0 \\ r_2 & r_2 & 0 & r_2 & \ldots & 0 & 0 \\ \vdots & \ddots & \ddots & \ddots & \ddots & \ddots & \vdots \\ 0 & \ldots & \ldots & \ldots & r_k & \ldots & 0 \\ \vdots & \ddots & \ddots & \ddots & \ddots & \ddots & \vdots \\ 0 & \ldots & \ldots & \ldots & r_n & r_n & r_n \end{pmatrix}^T \begin{pmatrix} \delta_1 \\ \delta_2 \\ \vdots \\ \delta_k \\ \vdots \\ \delta_n \end{pmatrix} = \begin{pmatrix} R_1 \\ R_2 \\ \vdots \\ R_k \\ \vdots \\ R_n, \end{pmatrix} \quad (3)$$

or for simplicity we write $\mathbf{r}^T \boldsymbol{\delta} = \mathbf{R}$, where subscript $T$ is the transpose operator. The solution vector $\mathbf{R}$ represents the test results, and thus known with some degree of test uncertainty, $\Delta R$, which we shall account for. On the other hand, while the matrix $\mathbf{r}$ is unknown, the distributions of the samples in the tubes is our choice, and that is simply the matrix $\mathbf{r}$ with nonzero elements replaced by ones,

$$\mathbf{I}_{i,j} = \begin{pmatrix} 1 & 1 & 1 & \ldots & \ldots & 0 & 0 \\ 1 & 1 & 0 & 1 & \ldots & 0 & 0 \\ \vdots & \ddots & \ddots & \ddots & \ddots & \ddots & \vdots \\ 0 & \ldots & \ldots & \ldots & 1 & 1 & 1 \end{pmatrix} \quad (4)$$

which is the distribution matrix ($n \times l$) that shows how samples from each patient (rows) are added to the tubes (columns). Finally, the vector $\boldsymbol{\delta}$ contains the information we seek, on positive and negative patients. Thus, our objective is find $\boldsymbol{\delta}$.

**Phase 1: Identifying immediate negatives.** The test results vector $\mathbf{R}$ may contain zero elements (i.e. tests that return negative). For each test tube $j$ that returns zero, i.e. $\mathbf{R}_j = 0$, all patients $i$ that have $\mathbf{I}_{i,j} = 1$ should test negative, otherwise the test tube would not have returned zero ($\mathbf{R}_j \neq 0$). Thus, at this phase we are able to identify $q$ patients that are negatives, with $0 \leq q \leq n$.

**Phase 2: Identifying all positives.** We rewire equation (3), excluding the negatives identified in the previous phase as $\tilde{\mathbf{r}}^T \tilde{\boldsymbol{\delta}} = \tilde{\mathbf{R}}$. Thus, all elements of $\tilde{\mathbf{R}}$ are now non zero. In order to identify all positives we seek all possible solutions for $\tilde{\boldsymbol{\delta}}$, using $(n-q)$ algebraic equations, for each possible solution $\tilde{\boldsymbol{\delta}}$. Now we carry out the following programme: (i) Consider all combinations that have a single positive, i.e. exactly $(n-q)$ combinations. Note that this can be the case only when elements in $\tilde{\mathbf{R}}$ are identical, within the test uncertainty $\Delta R$. (ii) Consider all combinations with two positives, which is also easy to identify/exclude as it requires $\tilde{\mathbf{R}}$ to have either two or three different elements; two numbers that each corresponds to one of the positive patients, and possibly a third in case they are combined in a third tube. (iii) In general, we keep add an additional possible positive (i.e. increase the number of 1's in $\tilde{\mathbf{R}}$), and seek for a solution.

**Phase 3: Identifying more negatives.** All patients that were not identified positive in phase 2, nor negative in phase 1, have to be negative, as all positives have been identified. Thus all patients with $i(\tilde{\boldsymbol{\delta}}_k = 0)$ are identified as negative.

## 2  Results & discussion

To gain more quantitative understanding of the proposed method, we performed numerical tests with actual scenarios under various conditions. We considered group sizes of 28, 56, and 70 patients that were tested using 8 tubes only (figure 1), and group sizes of 120, 210 and 252 patients using 10 tubes - see figure 2. Infected patients were selected randomly, with a percentage that ranged from 0.8 to 21.43. Each result point is an average of a hundred test repetitions, which is important when discussing uncertainties in the tests that were allowed to be between $\Delta R = 0.05\%...30\%$. Without loss of generality, each of the positively tested patients were allocated a random number between 0-220 nano grams (ng), that mimics possible reading, e.g. using a PCR technique, though any other range could be equally implemented.

Uncertainty in the test results, $\Delta R$, depends on a number of factors among which are the accuracy and precision of the test method. The less sensitive the test is the larger $\Delta R$ becomes, e.g. an uncertainty of 10% is equivalent to 20 ng so that we are unable to distinguish between two readings with difference that is less than 20 ng. Therefore, as $\Delta R$ increases false solutions may appear, and thus false positive results are expected. However, since the actual solution lies within any given uncertainty, we never obtain false negatives, which is extremely important for disease management and control. If the uncertainty is very small, we are always able to obtain a single solution $\boldsymbol{\delta}$ that determines accurately all positively and negatively tested patients. Unfortunately, current technology is associated with a degree of uncertainty that requires optimising the solution using other factors such as tube set size $m$, and subset $l$. The smaller the percentage of infected patients, the higher the percentage of negatives identified (figure 1). As the uncertainty increases, the size of subset $l$ becomes important. Having a smaller $l$ size results in a higher percentage of negatives, though the size of the group $n$ becomes smaller (figure 2 and eq.(1)). When $m$ and $l$ increase the length of $\tilde{\tilde{\delta}}$ becomes larger, which requires solving for a greater combination of possible solutions. If the uncertainty in the tests is large, finding a solution and determining the positives is not always possible, though if solutions are found we always determine all positives, even if there are false positives. For example, for 252,000 patients, we performed 100 independent tests with 10 tubes each, and allowed each patient samples to be distributed to 5 tubes (see right subplot of figure 2, blue-cross curve), even at high uncertainty of 32.77 ng (14.9% error), we were able to determine all positives in 90 tests out of 100 tests that were carried out. Thus, all positives from 226,800 patients were determined. In the remaining 8 tests, no solution was found and thus there is no risk for false negatives, in any of the tests. However, in this case we are only able to determine, on average 67.67% of all negatives in all 100 tests (see left subplot of figure 2, blue-cross curve). If a more accurate test is applicable, say with an uncertainty of 2.048 ng ($\simeq$ 1% error), %99.6 of all negatives are identified in all 100 tests. Note that in figure 2 the percentage of positives is variable and selected randomly to mimic an actual scenario were we don't know the percentage of infected patients a priori.

Although attention was focused here on enhancing lab tests for COVID-19, in particular, and epidemics in general, we believe that the proposed method can be equally implemented in a variety of lab tests. In fact, once the solutions vector $\boldsymbol{\delta}$ is obtained one could structure back quantities in each of the infected samples, i.e. $\mathbf{r} = \mathbf{R}^T \boldsymbol{\delta}$. Such a quantitative method could be complementary to a wide range of environmental, healthcare and safety, and engineering applications, e.g. for testing contaminated water or food, spread of diseases in population and sewage, and proper concentration of substances in chemical products.

**Methods**

**Numerical analysis** For the numerical calculations, we used a MacBook Pro, 2.8 Ghz Quad-Core Intel Core i7 processor, 16GB 1600 Mhz DDR3 memory. A code has been developed in Matlab R18a. An example of the output for a single test can be found in the supplementary information.

**Calculation algorithm**

- Each patient is allocated a binary number with $l$ one's and $(m-l)$ zero's, e.g. (0 0 1 0 1 1 0 ...). The location of each digit of the binary number corresponds to a single tube ($m$ tubes in total), e.g. the first digit from left (0) corresponds to the first tube, the second digit (0) corresponds to the second tube, the third digit (1) corresponds to the third tube, etc.

- Samples from each patient are added to all tubes that correspond to digits with number '1'. For example, samples from patient (0 0 0 1 1 1) are added to tubes number 4, 5, and 6 (from left).

- All $m$ tubes are sent for testing. It is important that testing is obtained at a cycle where no saturation has been reached, so that the value of each positive test represents a summation of unique values for each positive sample in that tube. A quantitative result ($\mathbf{R}^T$) is obtained in the following form:

  0   34.0972   159.3790   125.2818   159.3790   0   0   159.3790   ...

- All tubes that return negative indicate that all patients that have samples in that tubes are not infected (negatives).

- Given the test results and uncertainty within, we then calculate all possible combinations of solutions $\tilde{\boldsymbol{\delta}}$ that return $\tilde{\mathbf{R}}$, which when found, return all positives, and possibly additional negatives.

- In the unlikely scenario of no solution, a repeated test for all unidentified samples is required. This can be done by reshuffling samples, or splitting them between other test groups.

**Acknowledgements**   The author is grateful for M. Abu-Khalaf, A. Mansour, A. Kadri, and R. Asleh for fruitful discussions.

**Competing Interests**   The author declares that he has no competing financial interests.

**Correspondence**   Correspondence and requests for materials should be addressed to U.K. (email: kadriu@cardiff.ac.uk).

**Figure 1** **Fixed number of positives.** Uncertainty analysis in tests using 8 tubes ($m = 8$), for fixed number of randomly selected positives, ranging from 2.86% to 21.43% of total patients. Each calculation point is an average of 100 repetitions. Left column: percentage of negative patients successfully identified. Right column: solutions found for positive patients (note that once a solution is found 100% of positives are identified). Rows top to bottom: $l = 3$, 4, and 5, with $n = 28$, 56, and 70, respectively.

**Figure 2** **Random number of positives.** Uncertainty analysis in tests for random number of positives (randomly selected), ranging from 0.8% to 5% of total patients. Using 10 tubes ($m = 10$), and $l = 3$, 4, and 5, corresponding to a total number of patients $n = 120$, 210, and 252, respectively. Left plot: percentage of negative patients successfully identified. Right column: solutions found for positive patients (note that once a solution is found 100% of positives are identified).

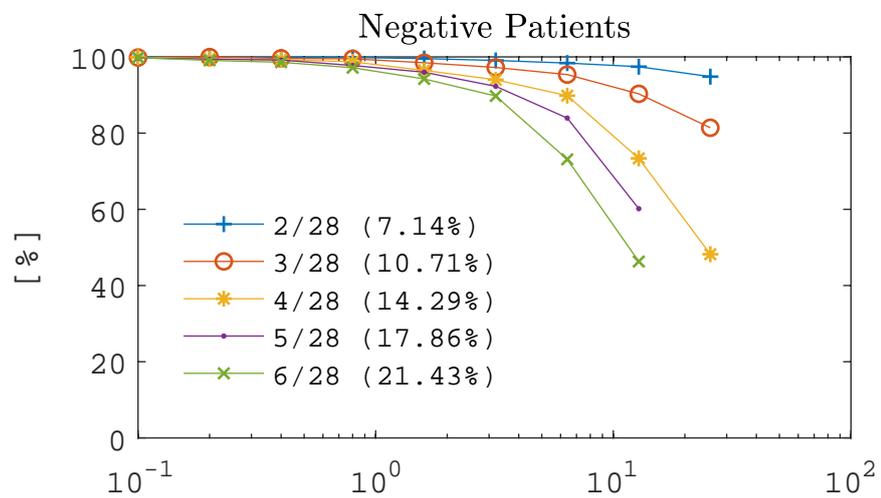
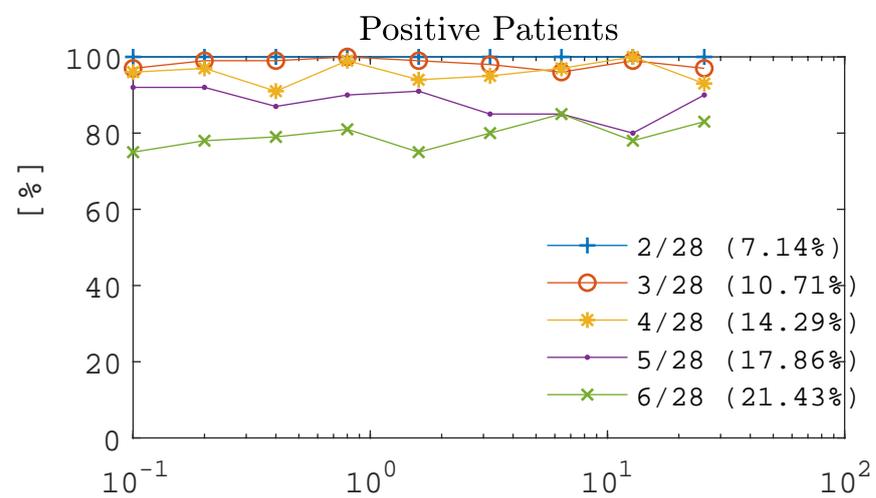
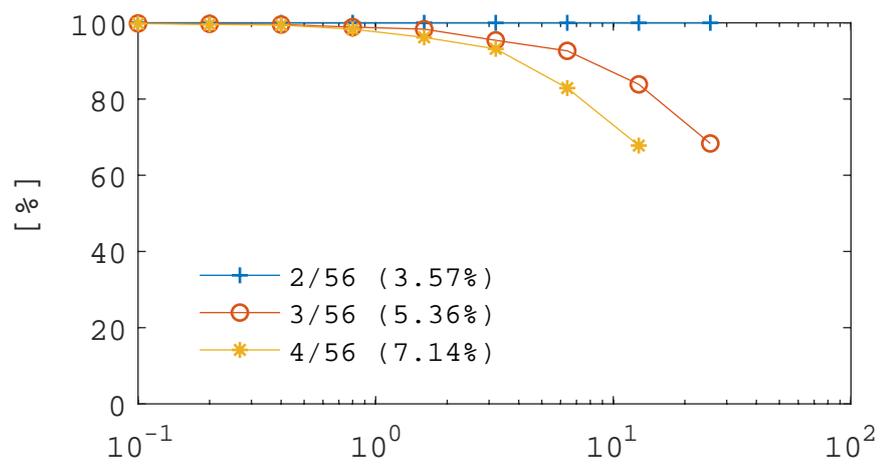
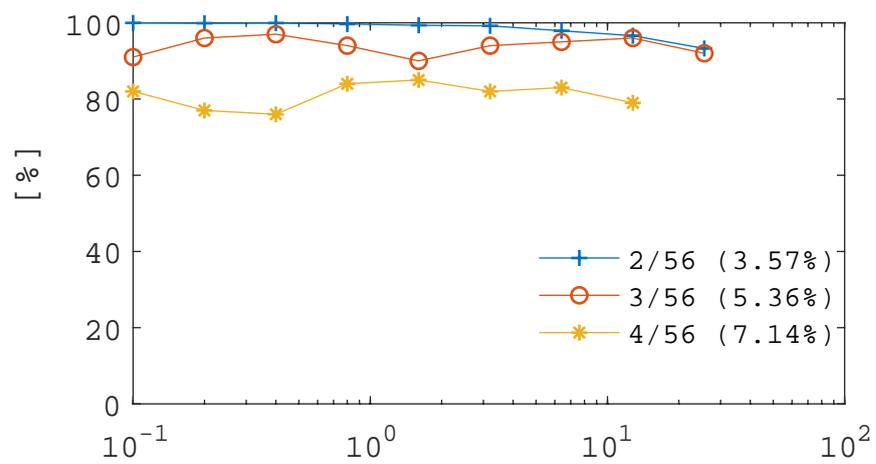
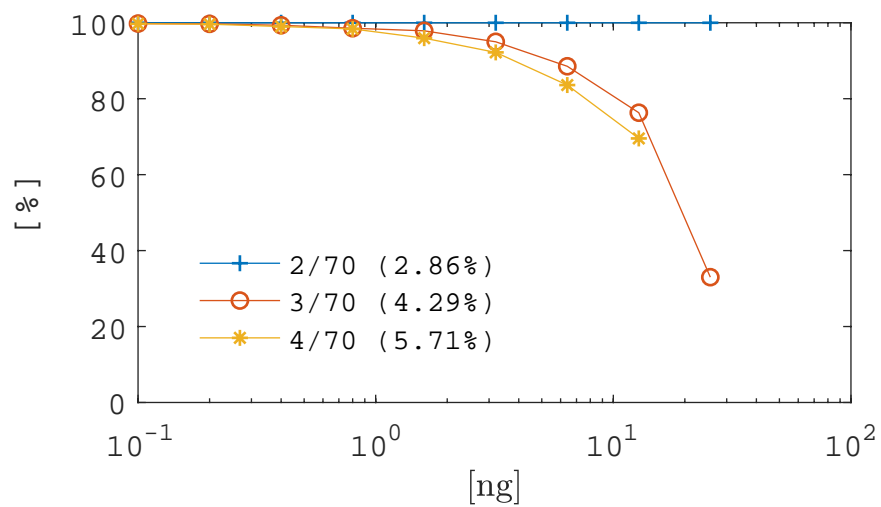
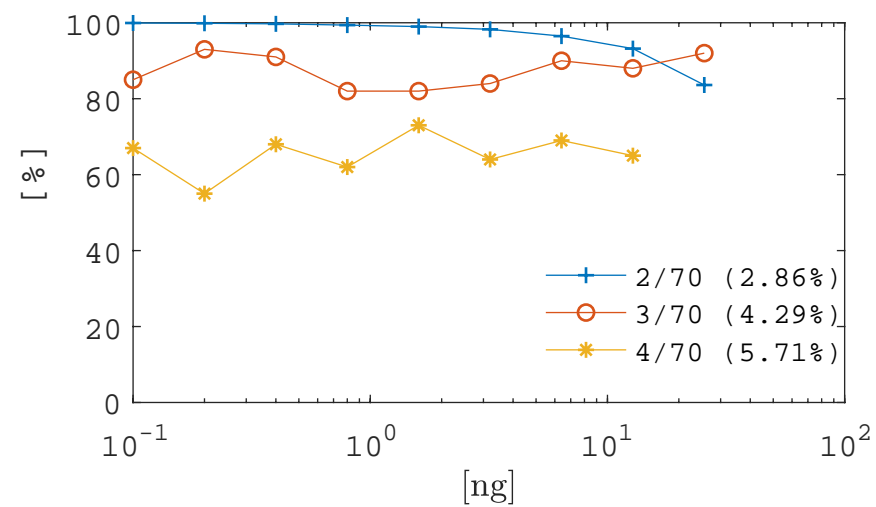

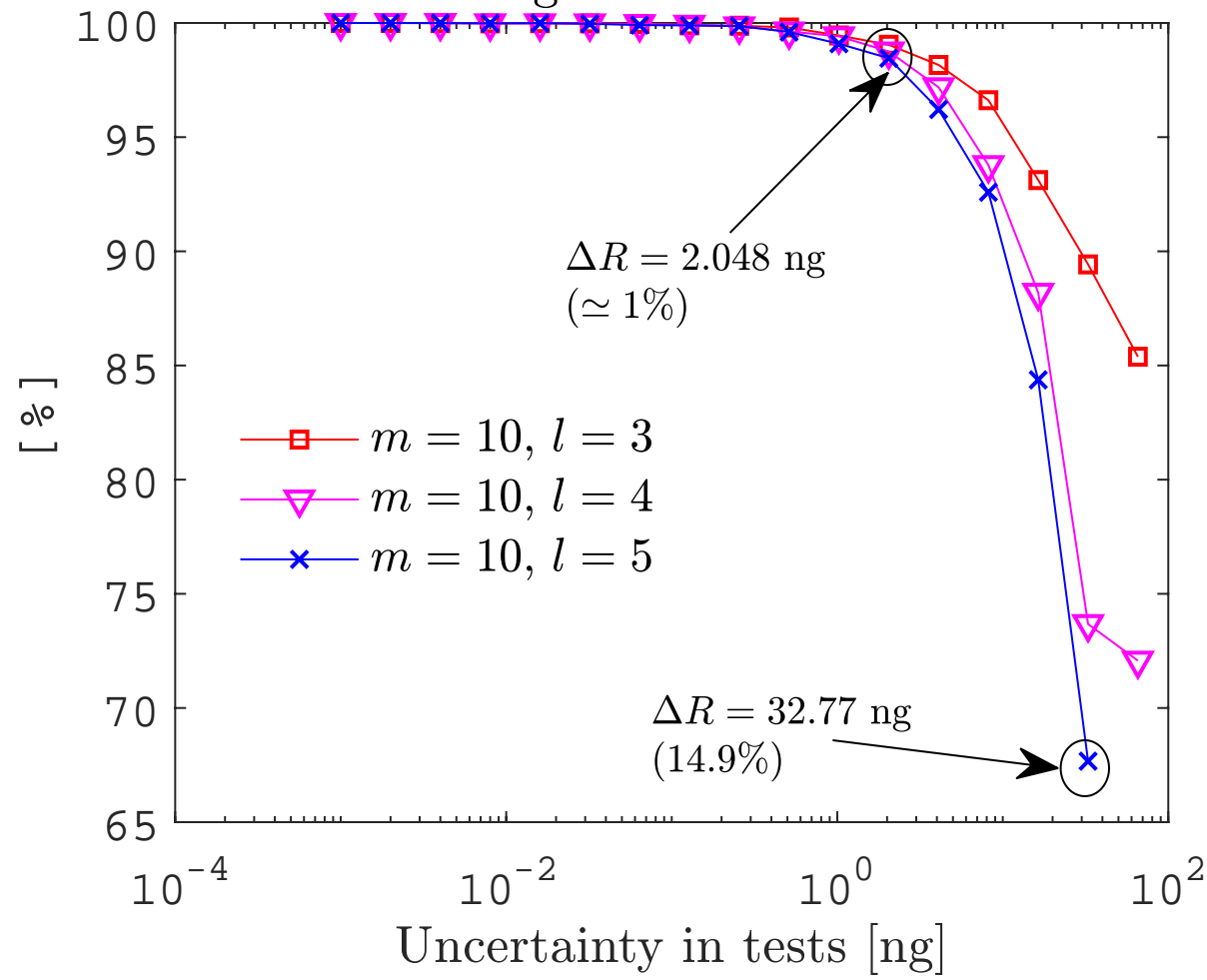 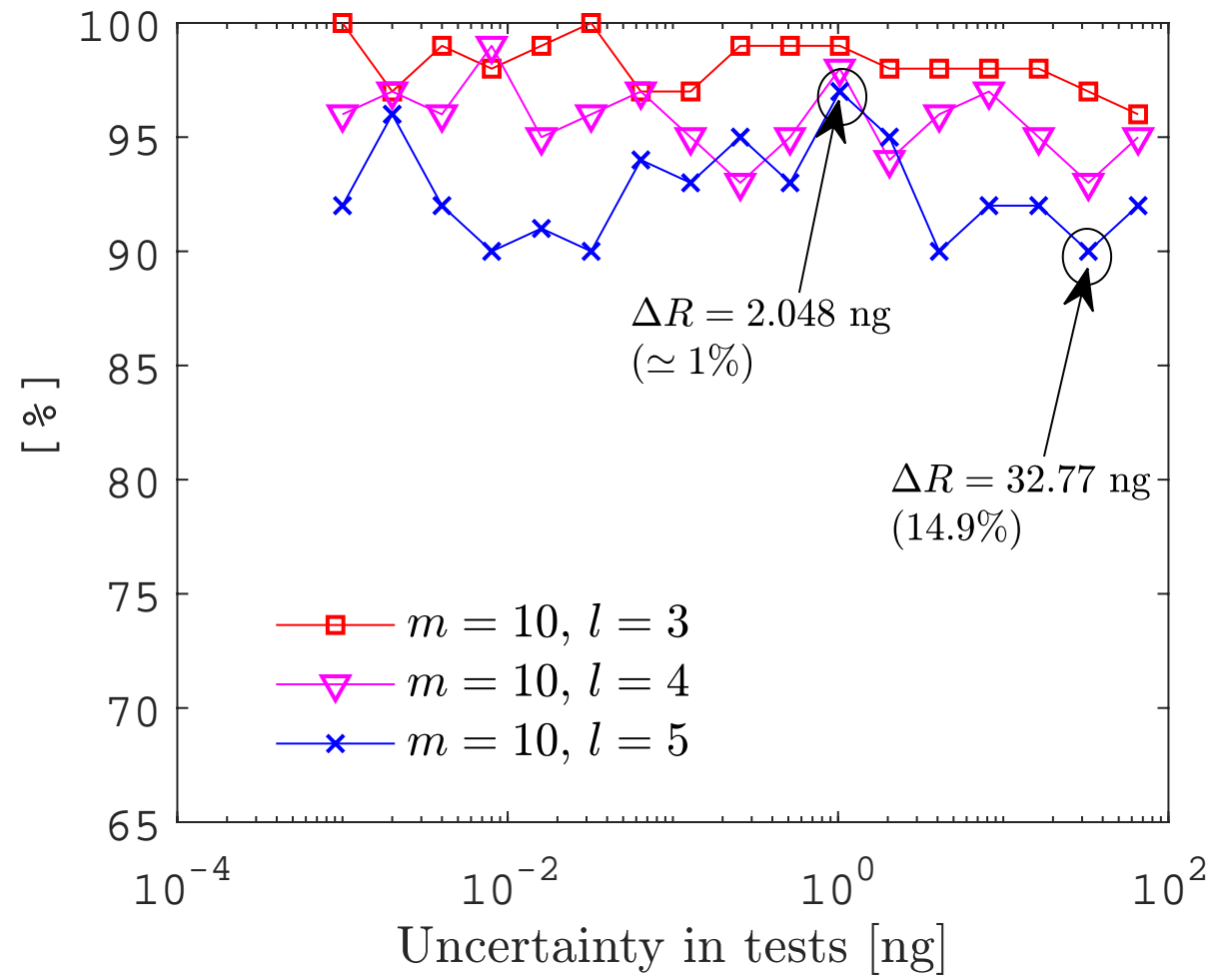